\documentclass[runningheads]{llncs}
\usepackage{graphicx}
\usepackage[utf8]{inputenc}
\usepackage{hyperref}\hypersetup{
  colorlinks=true,
  linkcolor=blue,
  filecolor=blue,      
  urlcolor=blue,
  citecolor=blue,
}
\usepackage{amsmath}
\usepackage[misc]{ifsym}
\usepackage{bbm}
\usepackage{cite}

\graphicspath{{picture/}}
\begin{document}
\title{DAGKT: Difficulty and Attempts Boosted Graph-based Knowledge Tracing}
\toctitle{DAGKT: Difficulty and Attempts Boosted Graph-based Knowledge Tracing}
%
%
\author{Rui Luo\inst{1} \and
Fei Liu\inst{1,2} \and
Wenhao Liang\inst{1} \and Yuhong Zhang\inst{1} \and \\Chenyang Bu\inst{1} (\Letter) \and
Xuegang Hu\inst{1} (\Letter)
\thanks{This work is supported by the National Natural Science Foundation of China (under grants 61806065, 62120106008, 62076085, and 61976077), and the Fundamental Research Funds for the Central Universities (under grants JZ2022HGTB0239).}
}
\tocauthor{Rui Luo, Fei Liu, Wenhao Liang, Yuhong Zhang, Chenyang Bu, Xuegang Hu }

\authorrunning{R Luo. et al.}
%
\institute{Key Laboratory of Knowledge Engineering with Big Data (the Ministry of Education of China), School of Computer Science and Information Engineering, Hefei University of Technology, China \and
Jianzai Tech, Hefei, China\\
\email{\{chenyangbu,jsjxhuxg\}@hfut.edu.cn}}
\maketitle   
\begin{abstract}
In the field of intelligent education, knowledge tracing (KT) has attracted increasing attention, which estimates and traces students' mastery of knowledge concepts to provide high-quality education. In KT, there are natural graph structures among questions and knowledge concepts so some studies explored the application of graph neural networks (GNNs) to improve the performance of the KT models which have not used graph structure. However, most of them ignored both the questions' difficulties and students’ attempts at questions. Actually, questions with the same knowledge concepts have different difficulties, and students' different attempts also represent different knowledge mastery. In this paper, we propose a difficulty and attempts boosted graph-based KT (DAGKT)\footnote{https://github.com/DMiC-Lab-HFUT/DAGKT}, using rich information from students’ records. Moreover, a novel method is designed to establish the question similarity relationship inspired by the F1 score. Extensive experiments on three real-world datasets demonstrate the effectiveness of the proposed DAGKT.\\ 
\end{abstract}

\section{Introduction}
In recent years, with the development of intelligent tutoring systems, more users choose online education because it is more convenient to provide personalized and high-quality education than traditional classrooms\cite{bu2022cognitive}. Knowledge tracing (KT), which evaluates students' knowledge mastery based on their performance on coursework, has attracted great attention and in-depth research.

Nowadays, KT models based on graph neural networks (GNNs) present satisfied performance, because there are natural graph structures among knowledge concepts (KCs) and questions in KT \cite{liu2021survey}. Nakagawa et al. \cite{2019Graph} proposed the graph-based KT (GKT) to learn the graph relations among KCs using the GNN. Graph-based interaction model for
KT (GIKT) \cite{2021GIKT} focuses on the relationships between questions and KCs, obtaining higher-order embeddings of questions and KCs by the graph convolutional network (GCN) \cite{GCN}. Question embeddings and answer embeddings in KT task \cite{piech2015deep} are integral parts of exercise embeddings. Among them, question and answer embeddings represent the information of questions and students' performance on questions, respectively.
These GNN-based KT models obtain satisfied performance because better exercise embeddings are achieved through question embeddings with graph relationships using GNNs.

\begin{figure}[t]
  \centering
  \includegraphics[scale=0.45]{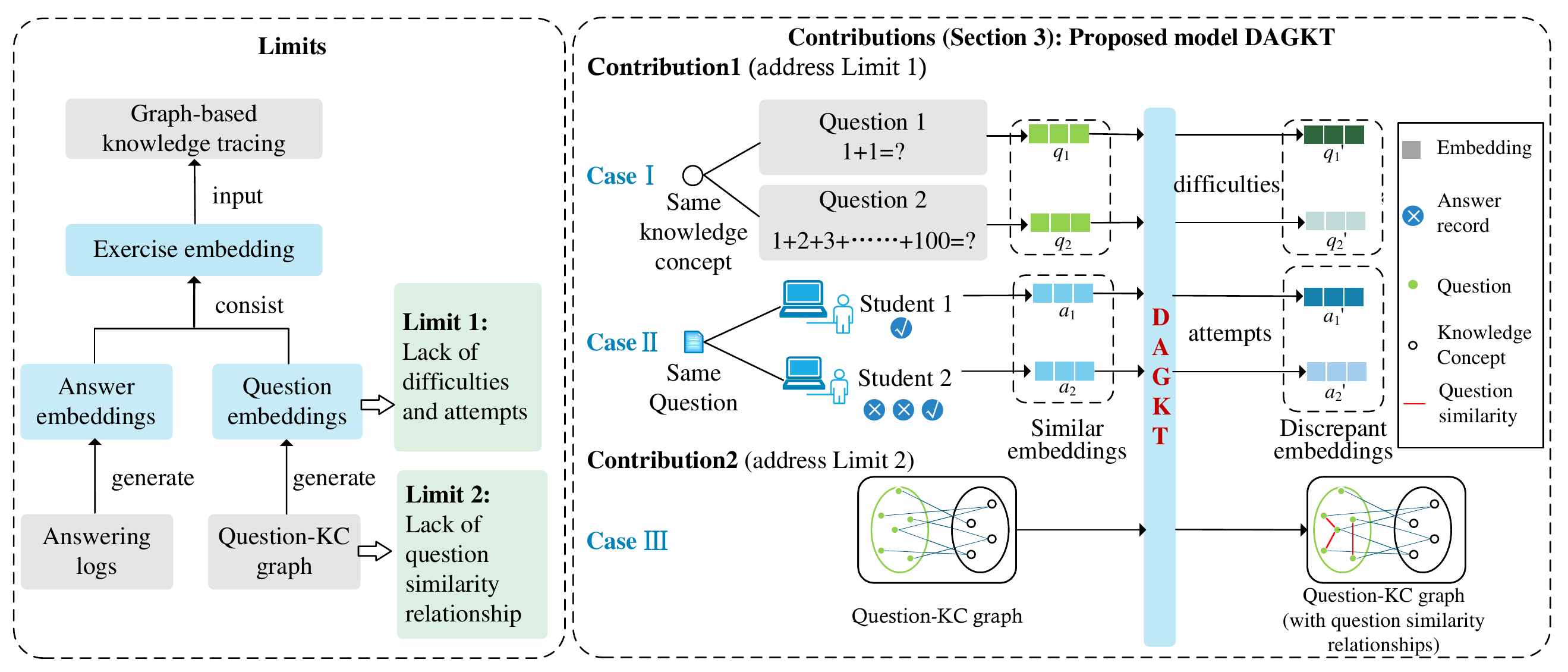}
  \caption{The limits and contributions. The limit of lack of difficulties and attempts is addressed shown in Case I and II, and the limit of lack of question similarity relationship is addressed shown in Case III.}
  \label{tab:fig1}
\end{figure}

Exercise embedding plays an important role in KT task, because cognition evaluation in KT relies on students' performance on exercises.
There is rich information involved in exercises such as stem texts \cite{xiao2021deep} and student behaviors features \cite{nagatani2019}. There is still room for improvement for both embeddings, analyzed from the following aspects, as shown in Case \uppercase\expandafter{\romannumeral 1}-\uppercase\expandafter{\romannumeral 3} of Figure \ref{tab:fig1}.

First, most existing GNN-based KT models ignore the question difficulties in question embeddings as well as attempts in answer embeddings. Difficulties and the number of attempts are critical as question embeddings and answer embeddings which reasons are as follows. When two questions $q_1$ and $q_2$ examine the same KC, student $s_1$ may give different answers because $q_1$ and $q_2$ have different difficulties (shown in Case \uppercase\expandafter{\romannumeral 1} of Figure \ref{tab:fig1}). And if the number of attempts is not considered, the model will think that student $s_2$  who has tried 10 times to get it right, and student $s_3$  who got it right after only one attempt have the same experience (shown in Case \uppercase\expandafter{\romannumeral 2} of Figure \ref{tab:fig1}). So if difficulties and attempts are not considered models can't discriminate between questions with the same KCs, or between answers with different attempts.

Furthermore, the question embedding  is achieved by GNN aggregating the information of the surrounding nodes in the question-KC graph, so the question-KC graph is very important. Most existing graph-based KT models perform convolution on bipartite graphs and there is no question-question relationship in the graph (shown in \uppercase\expandafter{\romannumeral 3} of Figure \ref{tab:fig1}). Gao et al. \cite{gao2021rcd} hold the view that there are two kinds of relationships between questions: prerequisite relationships and similarity relationships. In the field of GNN-based KT, few studies put the relationships between questions into the convolution process (most of them only use the question similarities in the prediction process, such as \cite{2021GIKT}).
Tong et al. \cite{tong2020hgkt} designed a method of constructing prior support relationships between questions from students' answer results illustrating the effectiveness of constructing relations from students' answer results.
However, most existing studies construct the question similarity relationship through question text information or problem embedding distances, without using the students' answer results. There is still a need for a method that can use  students' answer results to build similarity relationships.

To address these two problems, we propose the DAGKT model. Specifically, to solve the first problem, we design a fusion module to fuse two types of information: difficulty and attempts. We get the difficulties of the questions and the students' number of attempts from the datasets and encode them into embeddings through the encoder. After that, we put them with question embeddings and answer embeddings to the fusion module to obtain exercise embeddings that contain enormous information. Secondly, to address the second question and obtain a good question embedding, we design a relationship-building module that enriches the question-KC graph so that GCN can generate question embeddings that combine the information of the question relationships. We use statistical information combined with the calculation method of the F1 score to calculate similarity relationships between questions. It is assumed that the two questions may have a close relationship when students always obtain similar answering results (correct/incorrect) on the two questions. The F1 score is an indicator used in statistics to measure the accuracy of binary models. Another way to say, the F1 score infers to the degree of similarity between predicted and target values\cite{F1}. Therefore, the similarity of questions in this study is calculated according to the F1 score.

Finally, extensive experiments on real world datasets demonstrate the effectiveness of DAGKT and each module. In summary, our main contributions are as follows:

\begin{itemize}
    \item To address the problem that most graph-based KT models cannot clearly discriminate between questions with same KCs, or between answers with different attempts, DAGKT is proposed with a fusion module. In this module, the question and answer embeddings are fused with difficulty and attempts.

    \item  Furthermore, the relationship-building module is designed to construct the similarity relationship between questions, inspired by the F1 score. The constructed relationship enhances the representation of questions and improves the performance of KT.

    \item Several experiments are conducted on three public datasets. The results show that our model outperforms the baseline models and the effectiveness of the above two contributions is demonstrated.
\end{itemize}

\section{Related Work}
First, the KT models are reviewed. Then, the GNN-based KT model, i.e., GIKT, is introduced.
    \subsection{Knowledge Tracing}\label{KT}
KT is the task of estimating the dynamic changes in students' knowledge state based on their exercise records. Existing KT models can be categorized into two main types: Bayesian-based KT and deep learning KT models\cite{hu2020}. BKT is based on the hidden Markov model which is the first model proposed to solve the KT task. Several studies have integrated some other information into BKT, such as student’s prior knowledge \cite{pardos2010modeling}, slip and guess probabilities \cite{d2008more}, and student individualization \cite{yudelson2013individualized}. Due to the powerful ability to achieve non-linearity and feature extraction making it well suited to modeling the complex learning process,  deep neural networks have been leveraged in many KT models. DKT \cite{piech2015deep} uses a recurrent neural network to trace the knowledge state of the students which is the first deep KT model. DKVMN \cite{DKVMN} introduces an external memory module to store the KCs and accurately points out students’ specific knowledge state on KCs. Based on these two models, several studies consider adding more information to the models to improve their performance\cite{liu2021fuzzy}, such as the forgetting behavior of students \cite{nagatani2019} and student individualization \cite{minn2018deep}.\par
 With the development of GNN, it has been found that it can work well when dealing with graph-structured data. Nakagawa et al. \cite{2019Graph} presented the GKT, which uses GNNs to handle the complex graph-related data, such as knowledge concepts. GIKT \cite{2021GIKT} using GCN \cite{GCN} to obtain higher-order embeddings of questions through relations between questions and KCs. 
 
\subsection{GIKT}
In this subsection, we introduce the student state evolution module and prediction part in GIKT \cite{2021GIKT}. Our work is inspired by GIKT, an effective graph-based KT model, and we refer readers to the reference \cite{2021GIKT} for more details about GIKT.

\textbf{Student State Evolution Module:} For each time step $t$, GIKT concatenates the question and answer embeddings and transforms them into the representation of exercises through nonlinear layers:
\begin{equation}
    \mathbf{e}_{t}=\operatorname{ReLU}\left(\mathbf{W}_{1}\left(\left[\widetilde{\mathbf{q}}_{t}, \mathbf{a}_{t}\right]\right)+\mathbf{b}_{1}\right),
\end{equation}
where  $[,]$  denotes embedding concatenation. GIKT models the whole exercise process to capture the students' state changes and to learn the potential relationships between exercises. To model behaviors of students doing exercises, GIKT uses LSTM which can capture coarse-grained dependency like potential relationships between KCs to learn students' states from input exercise embeddings. And GIKT learns the hidden state as the current student state, which contains the coarse-grained mastery state of KCs.

\textbf{Prediction:}
To improve the performance of model, GIKT designed a history recap module that can select relevant history exercises (question-answer pair) to better represent a student’s ability on a specific question. GIKT chooses history questions sharing the same skills with the new question for prediction. After that, GIKT uses the interaction of cognitive state and questions, the interaction of cognitive states with related skills and interaction of the cognitive state at the time step of the relevant history exercise with the new question and its skills to predict, GIKT calculates the attention weights of all relevant interaction terms and computes the weighted sum as the prediction.

\section{The Proposed Model DAGKT}
In this section, our model is introduced in detail. We use the GIKT model as our base model because it is one of the state-of-the-art models in the graph-based KT field and can make good use of exercise embeddings to predict.
\subsection{Framework}
\begin{figure}[t]
  \centering
  \includegraphics[scale=0.34]{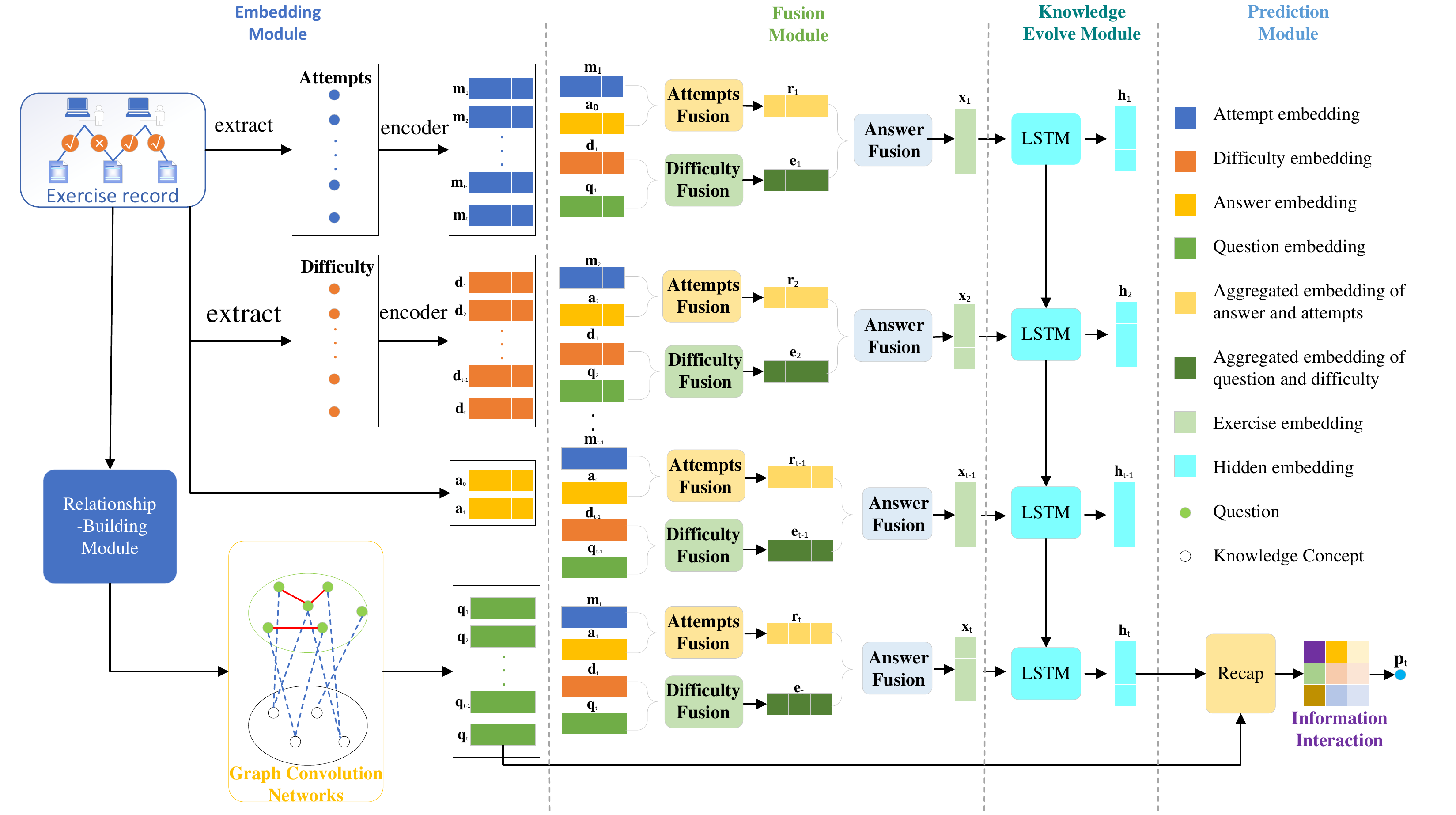}
  \caption{The framework of DAGKT, including four modules: embedding module (detailed in Section \ref{3.2}), fusion module (detailed in Section \ref{3.3}), knowledge evolution module (detailed in Section \ref{3.4}) and prediction module (detailed in Section \ref{3.4}).}
  \label{tab:fig2}
\end{figure}

The framework of DAGKT is shown in Figure \ref{tab:fig2}. First, we establish the similarity relationships between questions from the records to enrich the question-KC graph and generate the embeddings of questions through GCNs. Then, we extract the number of attempts students made on each question and the difficulty of each question from the records and encoder them into embeddings. After that, we put the question embedding, difficulty embedding, answer embedding and attempts embedding into the fusion module to obtain exercise embedding which denotes information about this exercise. Finally we put exercise embeddings into LSTM to obtain the knowledge mastery of students at each time step. And we make predictions through the prediction module.

\subsection{Embedding Module}\label{3.2}
In this subsection, we will introduce how the model produces three important parts of the four components of the exercise: question embedding, difficulty embedding and attempts embedding. The relationship-building module is used to generate similarity relationships between questions, and then the question embedding propagation is used to generate question embedding. Finally, attempts embeddings and difficulty embeddings are generated through the difficulty and attempts encoder module.\par
\textbf{Relationship-building Module:}
In this module, we introduce how to establish similarity relationships between questions. The same person's answers will be similar when the questions have similarity relationships. And the F1 score is a good indicator used in statistics to measure the performance of binary models which is used to measure how well the predicted results match the real results. Because the F1 score is a good indicator to measure the similarity of two sets of binary data, we use the true response result of the previous question as the prediction of the response result of the latter question, and calculate the F1 score between them as the similarity between them:
\begin{equation}
    \operatorname{Sim}\left(q_{1} , q_{2}\right)= \frac{(F_{1}\left(q_{1} , q_{2}\right)+ F_{1}\left(q_{2} , q_{1}\right))}{2},
\end{equation}
where $F_{1}\left(q_{1} , q_{2}\right)$ denotes when $q_1$ is answered before $q_2$, the similarity of the two answers. And $F_{1}\left(q_{1} , q_{2}\right)$ is calculated by:
\begin{equation}
    F_{1}\left(q_{1} , q_{2}\right)=2 \cdot \frac{P\left(q_{1}, q_{2}\right) \cdot R\left(q_{1}, q_{2}\right)}{P\left(q_{1}, q_{2}\right)+R\left(q_{1}, q_{2}\right)},
\end{equation}
where $P\left(q_{1}, q_{2}\right)$ and $R\left(q_{1}, q_{2}\right)$ are the parts used to calculate $ F_{1}\left(q_{1} , q_{2}\right)$. $P\left(q_{1}, q_{2}\right)$ and $R\left(q_{1}, q_{2}\right)$ are calculated by Eqs.(4) and (5).
\begin{equation}
    P\left(q_{1}, q_{2}\right)=\frac{\operatorname{Count}\left(\left(q_{1}, q_{2}\right)=(1,1)\right)+\lambda}{\sum_{a_{1}=0,1} \operatorname{Count}\left(\left(q_{1}, q_{2}\right)=\left(a_{1}, 1\right)\right)+\lambda},
\end{equation}
\begin{equation}
    R\left(q_{1}, q_{2}\right)=\frac{\operatorname{Count}\left(\left(q_{1}, q_{2}\right)=(1,1)\right)+\lambda}{\sum_{a_{2}=0,1} \operatorname{Count}\left(\left(q_{1}, q_{2}\right)=\left(1, a_{2}\right)\right)+\lambda},
\end{equation}
where $\operatorname{Count}\left(\left(q_{i}, q_{j}\right)=\left(a_{i}, a_{j}\right)\right)$ denotes the number of question sequences that reply $q_i$ with answer $a_i$ before $e_j$ with an answer $a_j$. Besides, to prevent the denominator from becoming too small, we introduced the laplacian smoothing parameter $\lambda = 0.01$ in Eqs. (4) and (5). After generating  similarity between the two questions, we add the edges between the questions $q_1, q_2$ to the question-knowledge concept graph when $\operatorname{Sim}\left(q_{1},q_{2}\right)$ is larger than hyperparameter $\omega$. So far, we have completed the construction of the question-KC graph.

\textbf{Question Embedding Propagation:}
We put the initialized question embeddings and question-KC graph into GCNs to obtain better question embeddings that have higher-order information between questions and KCs.

\textbf{Difficulty and Attempts Encoder Module:}
Since the difficulty and attempts play an important role in KT. In this module, we incorporate difficulties and attempts into exercise embeddings. Firstly, we obtain the number of attempts from the real dataset $M = \{m_{1,1},m_{1,2},\dots, m_{i,j}\}$ where $i$ denotes student ID and $j$ denotes question ID. Then we count the accuracy of all students on each question and calculate the questions' difficulties as
\begin{equation}
     d_i=\operatorname{function}(\frac{\operatorname{n}(correct(q_i))}{\operatorname{n}(correct(q_i))+\operatorname{n}(false(q_i))} ),
\end{equation}
where $\operatorname{n}(\cdot)$ denotes the number of $[\cdot]$.
After obtaining the attempts and difficulties, we use encoders such as three nonlinear fully connected layers to encode them into 100-dimensional embeddings $\mathbf{m}$ and $\mathbf{d}$:
\begin{equation}
     \mathbf{m_{i,j}} = \sigma (\sigma (\tanh(m_{i,j}))), \mathbf{d_{i}} = \sigma (\sigma (\tanh(d_i)))).
\end{equation}

And three more nonlinear layers are used to transform the 100-dimensional embeddings into numerical values:
\begin{equation}
     \tilde{m}_{i,j} = \sigma (\sigma (\tanh(\mathbf{m_{i,j}}))), \tilde{d_{i}} = \sigma (\sigma (\tanh(\mathbf{d_{i}}))).
\end{equation}
As the optimizer optimizing parameters, we obtain embeddings of attempts $\mathbf{m_{i,j}}$ and difficulties  $\mathbf{d_{i}}$ when the $\tilde{m}_{i,j}$ and $\tilde{d_{i}}$  come in close to the original inputs $m_{i,j}$ and $d_i$.

\subsection{Fusion Module}\label{3.3}
After embedding module, we can obtain question embedding $\mathbf{q}$, difficulty embedding $\mathbf{d}$, attempts embedding $\mathbf{m}$ and answer embedding $\mathbf{a}$. In this module, we fuse the four embeddings. First, we fuse the difficulties and questions through difficulty fusion module to obtain the aggregated embeddings of question and difficulty. Then we fuse the attempts and answers through attempts fusion module to obtain the aggregated embeddings of answers and attempts. Finally, the two aggregated embeddings are passed through one more layer of the nonlinear neural network to obtain the final exercise embeddings we need. The formula for the whole process can be expressed as follows:
\begin{equation}
    \mathbf{x}_{t} = \operatorname{Relu} (\mathbf{W}_{3}([\mathbf{W}_{1}([\mathbf{q}_{t}, \mathbf{d}_{t}])+\mathbf{b}_{1}, \mathbf{W}_{2}([\mathbf{a}_{t}, \mathbf{m}_{i,j}])+\mathbf{b}_{2}])+\mathbf{b}_{3}),
\end{equation}
where $\mathbf{W}_{1}, \mathbf{W}_{2}, \mathbf{W}_{3}$ and $\mathbf{b}_{1}, \mathbf{b}_{2}, \mathbf{b}_{3}$ are trainable matrices and parameters.

\subsection{Knowledge Evolution Module and Prediction Module}\label{3.4}
In this subsection, we describe how the model utilizes exercise embeddings to generate the cognitive state of students, how to use the cognitive state to generate predictions, and how to optimize the model.\par
\textbf{Knowledge Evolution Module:}
From the previous step, exercise embeddings absorb the relationships between questions and the relationships between questions and KCs, representing the behavior of students doing exercises. In each history step, to model the sequential behavior of students doing exercises, we put exercise embeddings into LSTM to learn the knowledge mastery changes of students where the hidden state denotes the current student state. 

\textbf{Prediction Module:}
When students do exercises, it is easy for them to associate the experience of doing similar questions to help them to solve the current questions. Like GIKT \cite{2021GIKT}, we select questions with similar KCs from historical questions to help the model make predictions. And we use the interaction of cognitive state and question, the interaction of cognitive state and related skills, and the interaction of the cognitive state at the time step of the relevant historical question with the current question to make predictions.

\textbf{Optimization:}
To optimize our model, we choose the method of Adam optimization, the parameters in the model can be updated by minimizing the loss function which contains three parts: (1) cross-entropy between the probabilities that the students will answer the question correctly $p_t$ and the true labels of the students’ answer $a_t$, (2) mean square error between difficulties before encoder $d_{i}$ and difficulties after decoder $\tilde{d_{i}}$ and (3) mean square error between attempts before encoder $m_{i,j}$ and attempts after decoder $\tilde{m}_{i,j}$:
\begin{equation}
\begin{split}
     \mathcal{L}=-\sum_{t}\left(a_{t} \log p_{t}+\left(1-a_{t}\right) \log \left(1-p_{t}\right)\right)\\+\sum_{n}\sum_{t}((m_{i,j}-\tilde{m}_{i,j})^{2})
        +\sum_{t}((d_{i}-\tilde{d_{i}})^{2}).
\end{split}
\end{equation}

\section{Experiments}
In this section, we conduct several experiments to investigate the performance of our model. We evaluate the prediction by comparing our model with other baselines on three public datasets. Then we make ablation studies show our modules' effectiveness in Section \ref{sec_abl}.
\subsection{Setup}
Datasets, baselines and implementation details are introduced in this subsection.\par
\textbf{Datasets:} To evaluate our model DAGKT, We have conducted extensive experiments on three public available datasets, i. e., ASSIST09 \footnote{https://sites.google.com/site/assistmentsdata/home/assistment-2009-2010-data/skill-builder-data-2009-2010.}, JUNYI\cite{junyi} \footnote{https://pslcdatashop.web.cmu.edu/DatasetInfo?datasetId=1198.} and CSEDM \footnote{https://pslcdatashop.web.cmu.edu/Files?datasetId=3458.}. In the ASSIST09 dataset, we remove the duplicated records and scaffolding problems. We randomly select 5000 students in JUNYI dataset as the whole dataset is too big\cite{2021GIKT}. Table \ref{tab:1} illustrates the statistics of the datasets.
\begin{table}[t]
\scriptsize
  \caption{Dataset statistics}\vspace*{1ex}
  \label{tab:1}
  \centering
  \begin{tabular}{ccccccc}
\hline
Numbers$\rightarrow$ & students & questions & skills & logs & questions/skills & skills/questions \\ \hline
ASSIST09 & 3,852      & 17,737      & 167      & 282,619     & 173                 & 1.12                \\
JUNYI    & 4,872      & 835         & 41       & 569,111     & 1                   & 20.36               \\
CSEDM    & 343        & 50          & 18       & 32,082      & 14.5                & 5.22                \\ \hline
\end{tabular}
\end{table}
We conduct a comprehensive examination of the models' performance in different fields and under different questions on  these three datasets: ASSIST09, which is a record of mathematical questions with both single and multiple knowledge concepts; JUNYI, which is a collection of mathematical questions with a single KC, and CSEDM, which is a collection of programming questions with multiple KCs. For each dataset, we take at least sequences with lengths greater than 3, as it is meaningless to be too short.

\textbf{Baselines:}
We select the following models as the baselines: 
DKT \cite{piech2015deep} \footnote{https://github.com/chrispiech/DeepKnowledgeTracing}, DKVMN \cite{DKVMN} \footnote{https://github.com/lucky7-code/DKVMN}, GKT \cite{2019Graph} \footnote{https://github.com/jhljx/GKT} and GIKT \cite{2021GIKT} \footnote{https://github.com/ApexEDM/GIKT} (detailed in Section \ref{KT}).

\textbf{Implementation Details: }
We initialize the 100-dimensional embeddings of KCs, questions. In the LSTM part, we use an LSTM with two hidden layers where the sizes of the memory cells are set to 200 and 100. For GCNs, we set the maximal aggregate layer number $L$ = 3. To avoid overfitting, we use a dropout with a keep probability of 0.8 for GCNs. We use the Adam optimizer to optimize the parameters of the model with the learning rate at 0.001 and the batch size of 32. We use Bayesian optimization to choose appropriate values for the other hyperparameters including  the number of related exercises to the new question, skills related to the new question, skill neighbors in GCNs and question neighbors in GCNs.\par
Five-fold cross-validation is used to obtain a stable experimental results, setting 80\% of the sequences as the training set and 20\% as the test set. We take the average of the best results for each fold as the final result. The comparison models use their own parameters, and each model is trained for 50 epochs. To evaluate the performance of each model on each dataset, we use the AUC as the evaluation metric, and the larger AUC the better performance of model.

\subsection{Overall Performance}
\begin{figure}[t]
  \centering
  \includegraphics[scale=0.45]{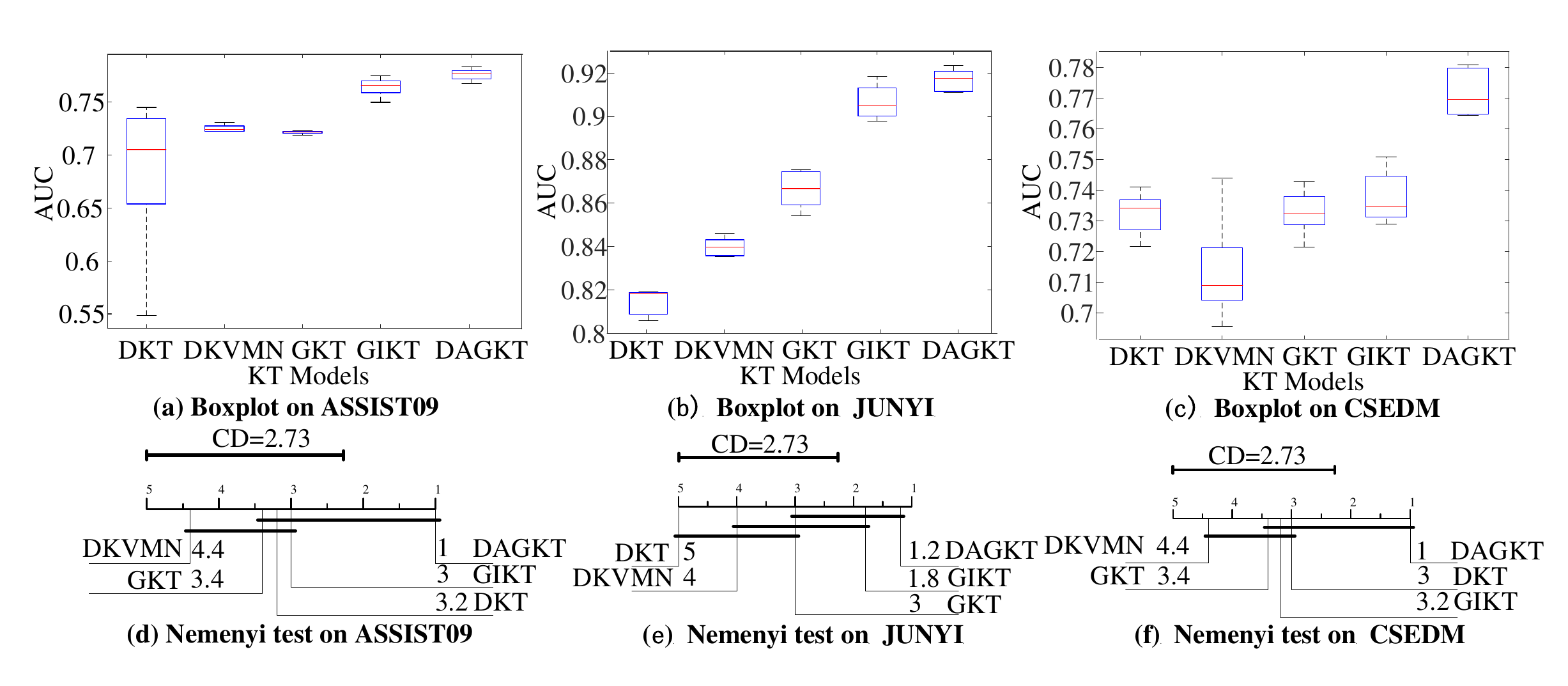}
  \caption{The boxplots and Nemenyi tests on three datasets. Our model performs best and the distribution is more concentrated.}
  \label{tab:fig3}
\end{figure}

Table \ref{tab:2} shows the AUC results of all the compared methods. Figure \ref{tab:fig3} (a-c) shows the boxplots on three datasets. From these results, it is found that our DAGKT model performs best on three datasets, which proves the effectiveness of our model. To be specific: (1) GKT, GIKT and DAGKT achieve better results than DKT and DKVMN, which shows the  effectiveness of GNN in handling the graph structure between question and KC. (2) The results of GIKT and DAGKT are better than those of GKT, which indicates that relationships of questions and KCs should be considered when constructing graph neural networks, and it is better to use relevant historical questions' information when predicting new questions. (3) DAGKT performs better than GIKT, which demonstrates the effectiveness of considering questions' difficulties, attempts and establishing similarity relationships between question. 
Moreover, the performance of our model is relatively concentrated, indicating that our model can obtain more stable results.\par
\begin{table}[t]
\scriptsize
\caption{Performance comparison on the datasets.}\vspace*{1ex}
  \label{tab:2}
  \centering
\setlength{\tabcolsep}{2.6mm}{
\begin{tabular}{cccccc}
\hline
Model$\rightarrow$ & DKT \cite{piech2015deep} & DKVMN \cite{DKVMN} & GKT \cite{2019Graph} & GIKT \cite{2021GIKT} & DAGKT \\ \hline
ASSIST09 &  0.6838   & 0.7253      &   0.7214  &   0.7641   &  \textbf{0.7759}     \\
JUNYI    & 0.8142    &  0.8398     & 0.8663     & 0.9066      &  \textbf{0.9168}      \\
CSEDM    &  0.7321   &  0.7138     &  0.7328   &  0.7378    & \textbf{0.7719}      \\ \hline
\end{tabular}}
\end{table}
To better illustrate the superiority of our model, we performed Nemenyi test \cite{Nemenyi} on three datasets, results as Figure \ref{tab:fig3} (d-f), the smaller the value, the better the performance of the model.

\subsection{Ablation Studies}
To demonstrate the effectiveness of the modules in DAGKT, we design the following models:

\begin{table}[t]
\scriptsize
\caption{Ablation studies of DAGKT. The relationship-building module, difficulty and attempts fusion module all improve model performance.}\vspace*{1ex}
  \label{tab:3}
  \centering
\begin{tabular}{ccccccc}
\hline
Model$\rightarrow$    & DAGKT-R & DAGKT-D & DAGKT-A & DAGKT-DA & DAGKT-G & DAGKT \\ \hline
ASSIST09 & 0.76414 &  0.7690&0.77146 & 0.7729    & 0.7647        &  \textbf{0.7759}     \\
JUNYI   &  0.9066    &   0.9099      &   0.9079      &    0.9135      &    0.9106     &    \textbf{0.9168}   \\
CSEDM    & 0.7378 & 0.7570 & 0.7560   & 0.7688  & 0.7577    &    \textbf{0.7719}   \\ \hline
\end{tabular}
\end{table}
\begin{itemize}
    \item DAGKT-R {removes relationship-building module and not fuse difficulties and attempts into exercise embeddings.}
    \item DAGKT-D  {only adds difficulty information to help the model predict.}
    \item DAGKT-A {only adds attempts information to help the model predict.}
    \item DAGKT-DA {adds information on both difficulty and number of attempts to help model predict.}
    \item DAGKT-G {uses the relationship-building module to construct similarity relationships between questions so that it can produce better exercise embeddings to help model predict.}
\end{itemize}
Table \ref{tab:3} shows that DAGKT achieve best performance considering similarity relationships between questions, difficulty of the question and the number of attempts. The results of DAGKT-D and DAGKT-A are better than  DAGKT-R, which shows the effectiveness of fusing difficulties and attempts. The results of DAGKT-D and DAGKT-A are better than  DAGKT-R, which shows that the attempts and difficulty have different improvement effects and the improvement of the effects are more stable when used at the same time. The results of DAGKT-G are better than DAGKT-R, which shows the effectiveness of relationship-building module which constructs similarity relationships between questions.
\label{sec_abl}

\section{Conclusion}

In this paper, to solve the problem that most existing graph-based KT models do not consider difficulty and attempts and do not establish similarity relationships between questions, we propose the DAGKT model which digs into questions' difficulties and number of attempts. Moreover, we design the relationship-building module to calculate the similarity between questions through the F1 score calculation method to establish similarity relationships between questions. Several experiments on three datasets demonstrate the effectiveness of the proposed model and the modules in the model.















\end{document}